\documentstyle[pra,aps,twocolumn]{revtex}

\begin{document}
\preprint{DTP 96--57\hskip1cm cond-mat/9612019}
\draft
\title{Interaction dependence of composite fermion effective masses}

\author{Uwe Girlich\thanks{e-mail: 
{\tt Uwe.Girlich@itp.uni-leipzig.de}}}

\address{Institut f\"ur Theoretische Physik,
Universit\"at Leipzig,
Augustusplatz 10,
D-04109 Leipzig,Germany}

\author{Meik Hellmund\thanks{e-mail: 
{\tt Meik.Hellmund@durham.ac.uk}}\thanks{
permanent address: Institut
f\"ur Theoretische Physik, Augustusplatz, Leipzig}}

\address{Department of Mathematical Sciences,
South Road, Durham DH1 3LE, England}
\date{\today}
\maketitle

\begin{abstract}
 We estimate the composite fermion effective mass 
for a general two particle potential  $r^{-\alpha}$
using exact diagonalization 
for polarized electrons in the lowest Landau level
on a sphere.
Our data for the ground state energy at
filling fraction $\nu=1/2$ as well as estimates of the excitation gap
at $\nu=1/3,\,2/5$ and $3/7$ show that $m_{\text{eff}} \sim \alpha^{-1}$. 
\end{abstract}
\narrowtext

The dynamics of interacting planar electrons in the lowest Landau
level of a strong magnetic field show many interesting features
at filling fractions $\nu<1$ experimentally observed as the fractional
quantum Hall effect (FQHE). It emerged that the picture of composite
fermions (CF)\cite{Jai89a,LF91}
 moving in a reduced magnetic field is central to the
understanding of the FQHE\@.  The field theoretic formulation of this
idea has received much attention (see e.g. \cite{SH95,Khv95}), in
particular  after Halperin, Lee and Read\cite{HLR93}
 described the
polarized $\nu=1/2$ state as a fermi liquid state of composite fermions.
Since the CF picture explains 
gaps due to the electron--electron interaction  as Landau level gaps
of composite fermions, their effective mass has to be understood 
as a result of this interaction.  

Numerical diagonalization of the interaction Hamiltonian for 
electrons on a sphere has a long history\cite{Hald83} as a testing
ground for the understanding of the FQHE\@. Rezayi and Read\cite{RR94}
have shown that even for the small number of electrons $N \approx  10$
accessible to exact diagonalization the pattern of the angular momenta
of $\nu=1/2$  ground states follows Hund's  rule
applied to composite fermions in a zero magnetic field. Later on, Morf and
d'Ambrumenil\cite{MdA95}   demonstrated that the ground state energy
itself allows an interpretation in the CF language and they estimated the CF
effective mass. This effective mass is also
the relevant parameter\cite{HLR93} 
for the excitation gap of $\nu=p/(2p+1)$ FQHE
states\cite{DM89}. 
The basic features of the FQHE as seen in
finite size studies are to a high degree independent of the exact 
form of the
two particle interaction potential $V(r)$. Most studies 
therefore used a simple $1/r$ potential. 
With the advent of a qualitative theory of the FQHE it now  seems
appropriate  to study the dependence of the 
numerically obtained effective masses
on the chosen potential. 

The single particle wave functions on a sphere of radius $R$ 
pierced by $\Phi=2S$ flux quanta are monopole harmonics
of angular momenta $j=S, S+1,\dots$ with energy
\begin{equation}
  \label{EE}
E=\frac{\hbar^2}{2m R^2} [j(j+1)-S^2].  
\end{equation}
We will use the ion disc radius $a=(\pi\,\text{density})^{-1/2}$ 
as basic 
length unit:  $R=a\sqrt{S}$.   
It is related to the magnetic length by $a=2l_0\sqrt{S/N}$. 

The quasipotential
coefficients\cite{Hald83}
for an interaction potential 
$V(r)= \frac{e^2}{\varepsilon  a} \left(\frac{a^{}}{r}\right)^\alpha$ 
with chord distance $r$ are 
\begin{eqnarray}
  V_J &=&(-1)^{2S+J}\frac{(2S+1)^2}{N^{\alpha/2}} 
 \frac{ \Gamma(1-\frac{\alpha}{2})}{\Gamma(\frac{\alpha}{2})}\nonumber\\ 
&&\times\sum_k (2k+1)\frac{ \Gamma(\frac{\alpha}{2}+k)}
{\Gamma(2+k-\frac{\alpha}{2})}
 \left\{ {S\atop S}{S\atop S}{k\atop J}
\right\}_{\text{6j}} 
\left({S\atop -S}{S\atop S}{k\atop 0} \right)^2_{\text{3j}}.
\end{eqnarray}
Fig.~\ref{FigQ} gives an impression of the $\alpha$ dependence of the $V_J$.
The lowest eigenvalues and eigenvectors of the Hamiltonian matrix have
been computed by a conjugate gradient algorithm\cite{bunk}.

We have calculated the ground state energy and angular momenta for 
$\nu=1/2$ systems with up to $N=13$ electrons.
The composite fermions feel no magnetic field at this filling fraction
and form a ``CF-atom'' with shells $j=0,1,2,\dots$ of degeneracy 
$2j+1$. 
The ground state angular momentum follows Hund's rule applied
to the CF-atom (e.g. $J=0$ for $N=n^2$ indicating $n$ closed shells) 
for $0.2\leq\alpha\leq1.99$. At $\alpha=0.1$ we find small
derivations in two cases
($L=1$ instead of 3 for $N=6$ and $L=4$ instead of 6 for
$N=12$) but these ground states are almost degenerate with states with the
``right'' angular momentum.  

Morf and d'Ambrumenil\cite{MdA95} found that the ground state energy 
per particle of the
$\nu=1/2$ system with Coulomb interaction can be interpreted, up to a
correction linear in $1/N$, as kinetic energy $T(N,m^*)$ of the CF-atom
with effective CF mass $m^*$.
This energy is calculated by summing up the contributions of the
individual particles given by eq.~(\ref{EE}) with $S=0$ and $m=m^*$.
It can be written as sum  of a part linear in $1/N$ 
(in units of $a$) 
and a  part which  vanishes for closed shells.
This deviation of the energy of partially filled  shells from 
linear behaviour is proportional to 
the effective mass parameter
$C$ introduced in \cite{HLR93}
\begin{equation}
  \label{C}
  \frac{\hbar^2}{m^* a} = \frac{C} {2}\frac{e^2}{\varepsilon}. 
\end{equation}
We find the same pattern of $N$-dependence of the ground state
energies in the range $0.1 \leq\alpha\leq1.99$. Fig.~\ref{Figfit}(a) 
shows that for
a long-range potential $\alpha=0.1$ the ground state energy comes very  
close to the  prediction of free composite fermions. 

On the other side, composite fermions are not free.  The interaction 
energy of particles in shells describes a similar pattern with 
relative minima for closed shells. 
In order to test the influence of CF interactions on this
method of obtaining $m^*$ we assume that the composite fermions  
interact via the same potential $r^{-\alpha}$ as the electrons.
The energy of closed shells as well as the inter--shell energy can
be calculated analytically  using the shell
model formalism of nuclear theory\cite{Talmi}.
 The ground state energy of the outer
partially filled shell is calculated numerically by exact diagonalization.
The sum of these contributions is
$V(N,\alpha)$, the ground
state energy of $N$ interacting particles of infinite mass on a sphere
without magnetic field.
We then checked that the electron ground state energies can be 
fitted by $a_0+a_1/N+V(N,\alpha)+T(N,\bar m^*)$. 
This provides another value for the effective mass $\bar m^*$,
calculated for {\em interacting\/} composite fermions.
For $\alpha<1$ (see Fig.~\ref{Figfit}(a)) this appears to be less
convincing than the free CF ansatz suggesting that in this case the CF
interaction is even weaker.
For $\alpha>1$, however, Fig.~\ref{Figfit}(b)) 
shows that the data can be interpreted by assuming a larger 
effective mass of the interacting composite fermions. 

The resulting values for $C$ and $\bar C$, as shown in 
Fig.~\ref{FigGapfit},
are (for $\alpha$ not too big) linear in $\alpha$:  
  $\bar C= 0.164\alpha,\quad C = 0.195\alpha$.

The Coulomb system at filling fractions $\nu=p/(2p+1)$
has been studied extensively  in the past, cf.\ e.g. \cite{KWJ96,DM89}.
Apart from the ground state at $L=0$ one finds a band of low-lying
excitations with $L=1,2,\dots,\Phi^*+1$, the exciton or magnetoroton.
($\Phi^*$ is the reduced magnetic field of the CF picture.) 
In the CF picture the ground state corresponds to $p$ filled Landau
Levels and the excitation gap in the limit $L\to\infty$ measures the distance
to the ($p+1$)th level. 
Fig.~\ref{FigExcit} shows the $\nu=1/3$ exciton mode, getting flatter for 
long-range potentials but with a visible magnetoroton minimum  at
$Ll'_0/R\approx1.4$. 

At $\nu=3/7$ the case $N=12$  is the only one
accessible to numerical diagonalization. $(N=9, \Phi=16)$  for
example is also
a $\nu=1/2$ state (an effect called ``aliasing''
in\cite{DM89}) and since the reduced magnetic field is zero, this
seems to be the preferable interpretation.
This makes a systematic study of finite size effects
impossible.  
Therefore we take the gap $\Delta E$ at $L=\Phi^*+1$ 
for the highest available electron
number ($N=10$ for $\nu=1/3, 2/5;$ $N=12$ for $\nu=3/7$) as an estimate   
for the CF gap between the $p$ and ($p+1$)th Landau level. 
The gap energies  for different $\nu$ 
fit to one curve if divided by $\Delta\nu=1/2-\nu$ (Fig.~\ref{FigEGap}). 
The CF picture expects a large $p$ behaviour 
\begin{equation}
  \Delta E=\frac{C}{p} \frac{e^2}{\varepsilon a}.
\end{equation}
 With $\Delta\nu  \to  \frac{1}{4p}$ as $p\to  \infty$
 our fit gives  $C = 0.35 \alpha$. 
We therefore  find an enhancement of $m^*$ for all potentials
by roughly a factor of $2$ as
$\nu\to1/2$.
d'Ambrumenil and Morf report\cite{DM89} that for Coulomb interaction  
a correction of the gap energy by the  
potential energy of two charges $q=1/(2p+1)$ at distance $2R$ allows a
consistent extrapolation in $1/N$.
Unfortunately, this does no longer work for systems 
with $\alpha  \neq1$. Nevertheless, our data are consistent  
with the value $C\approx  0.31$ for $\alpha=1$ extracted from
their data in \cite{HLR93}.    
However, the dependence of $E$ on $\Delta\nu$ appears to be linear 
for all values of $\alpha$ contrary to the analytical result 
$\Delta E\sim |\Delta\nu|^{\frac{2+\alpha}{3}}$ of \cite{Khv95}.

MH is supported by  
Deutscher Akademischer Austauschdienst.


\newpage
\begin{figure}
\caption{The first three quasipotential coefficients $V_i$ for
    $N=12, \nu=1/2$ as functions of $\alpha$. }
\label{FigQ}
\end{figure}

\begin{figure}
\caption{The ground state energy per particle plotted as function of
  $1/N$. The part of the energy linear in $1/N$ is subtracted.
For the data  shown by stars $V(N,\alpha)$ was first subtracted. 
(a) Data for $\alpha=0.1$; (b) data for $\alpha=1.5$.  }
\label{Figfit} 
\end{figure}

\begin{figure}
\caption{The mass parameters $C$ and $\bar C$ as functions of $\alpha$. The
  fits shown are $C=0.195\alpha$ and $\bar C=0.164\alpha$.}
\label{FigGapfit} 
\end{figure}

\begin{figure}
\caption{Exciton mode at $\nu=1/3$ with data from $N=8,9,10$ electron
  systems as function of the wave vector $L l'_0/R$.  
  The energy is in units of $e^2/\varepsilon l'_0$ with 
$l'_0$ the ``corrected magnetic
  length'' of Ref.~\protect\cite{DM89}, 
  $l'_0=a\sqrt{\nu/2}$. The lines are a
  guide to the eye only.}
\label{FigExcit} 
\end{figure}

\begin{figure}
\caption{Exciton gap energies in units of $e^2/(\varepsilon a\Delta\nu)$
     for $\nu=1/3, N=10;$ $\nu=2/5, N=10; \nu=3/7,
    N=12$.  The shown fit is $E=1.40 \Delta\nu  \alpha$.}
\label{FigEGap}
\end{figure}

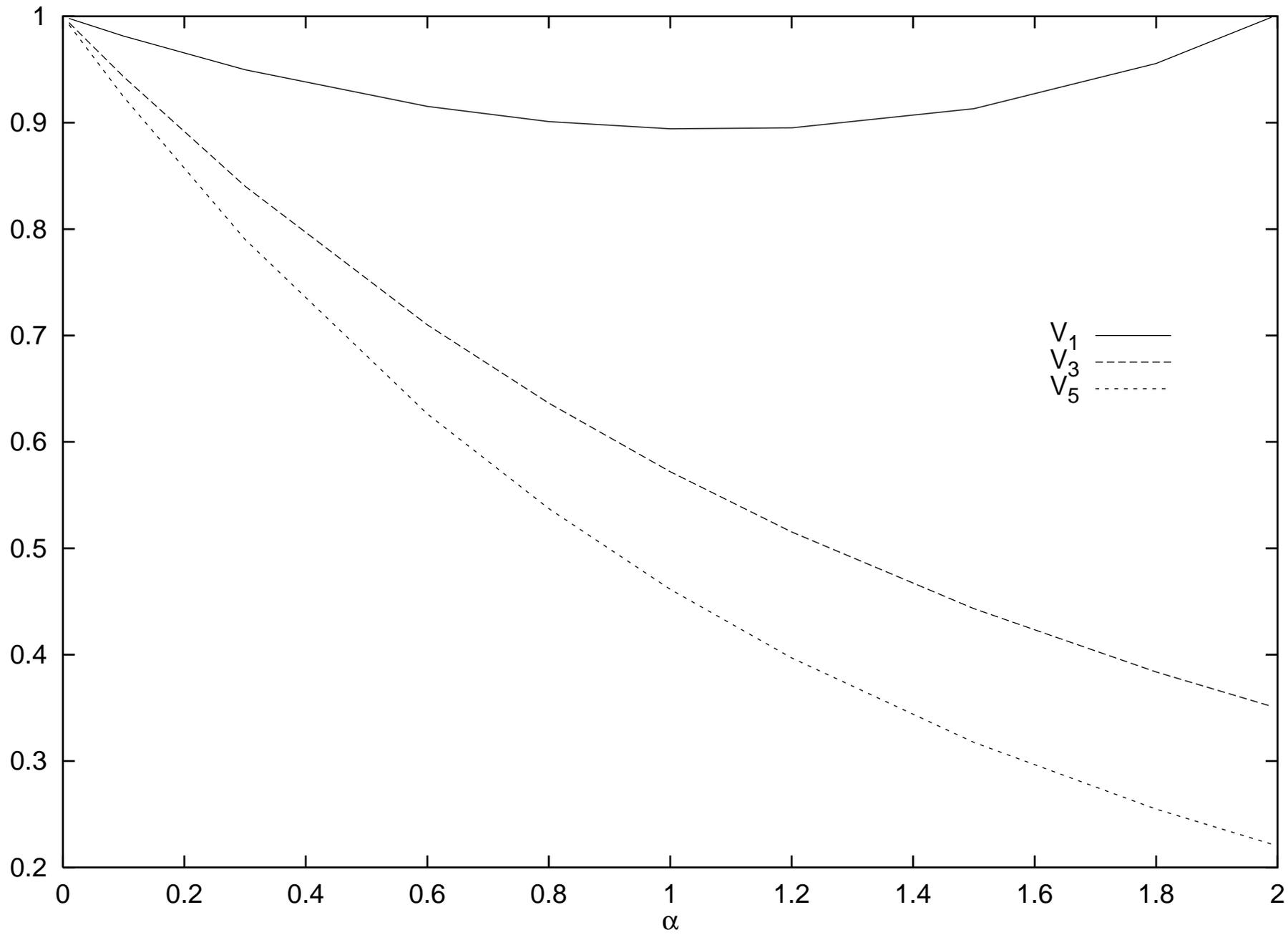
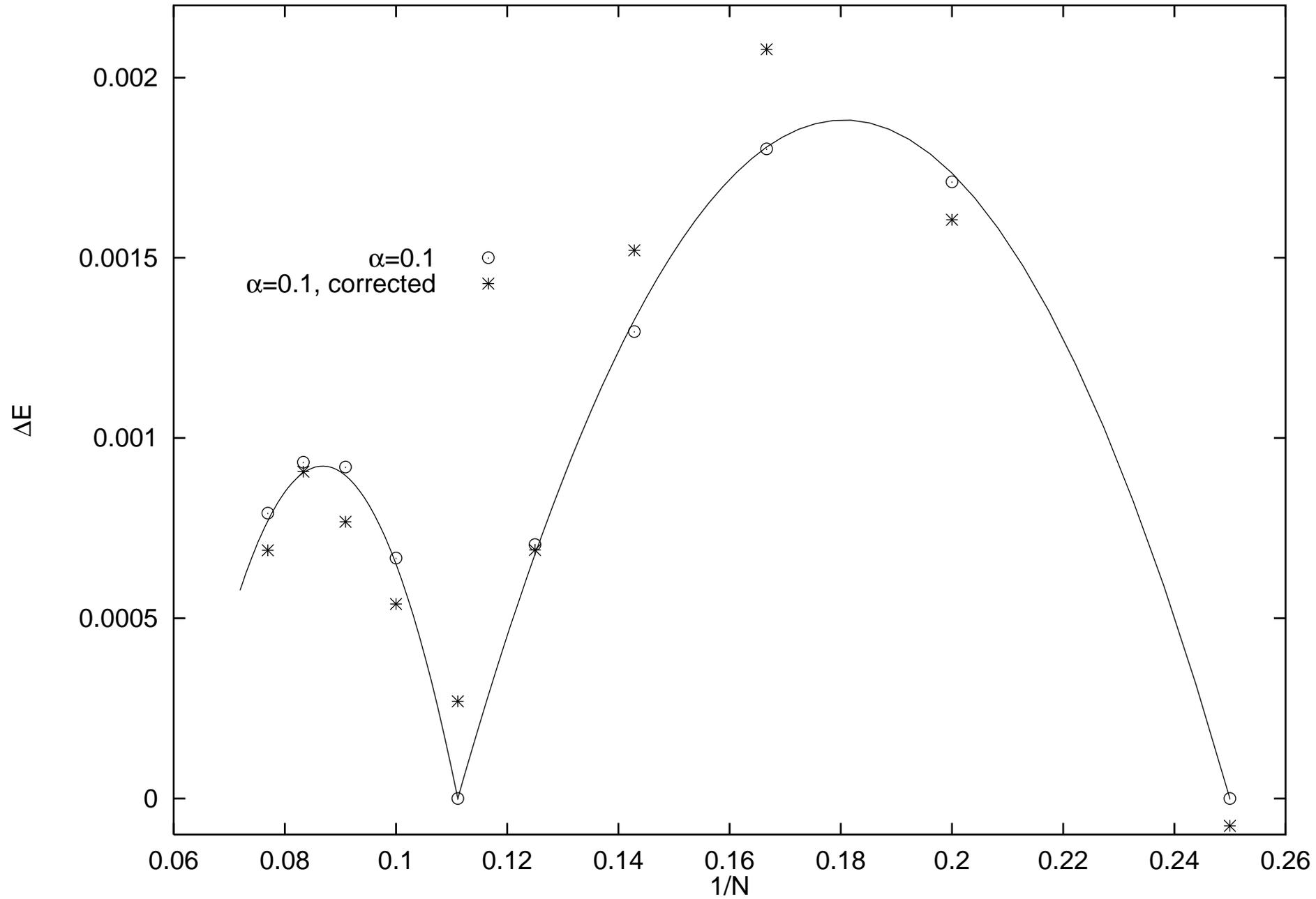
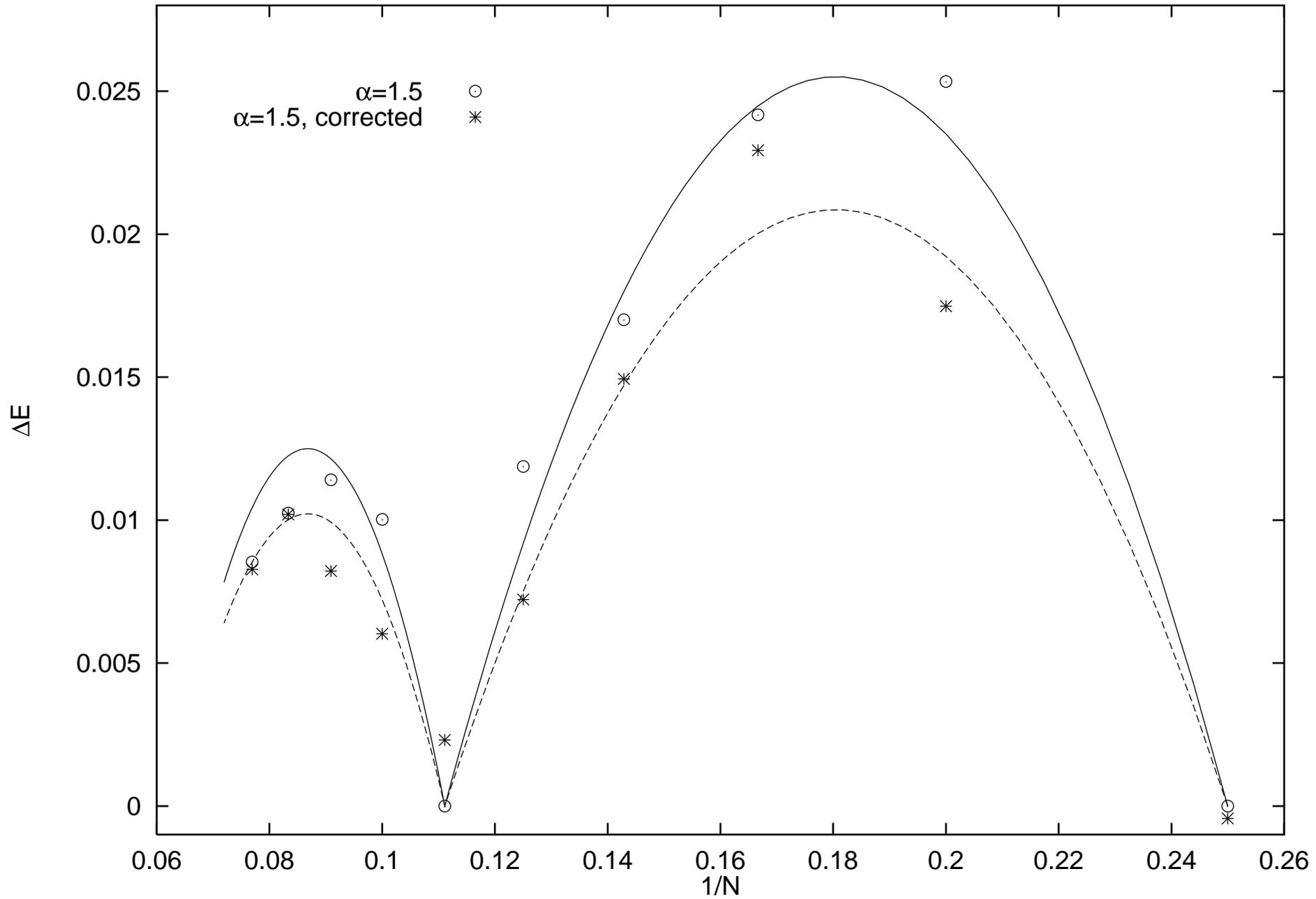
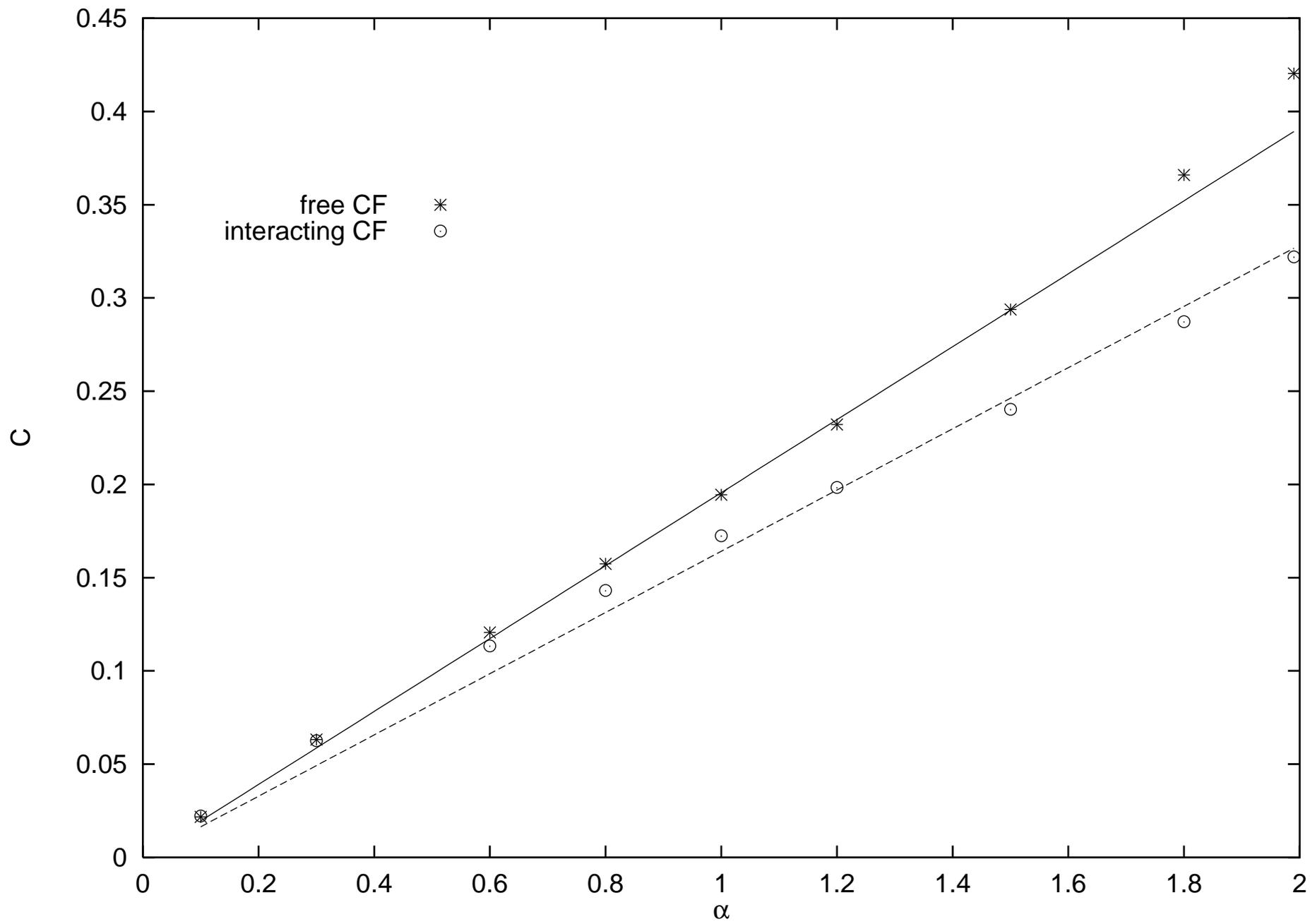
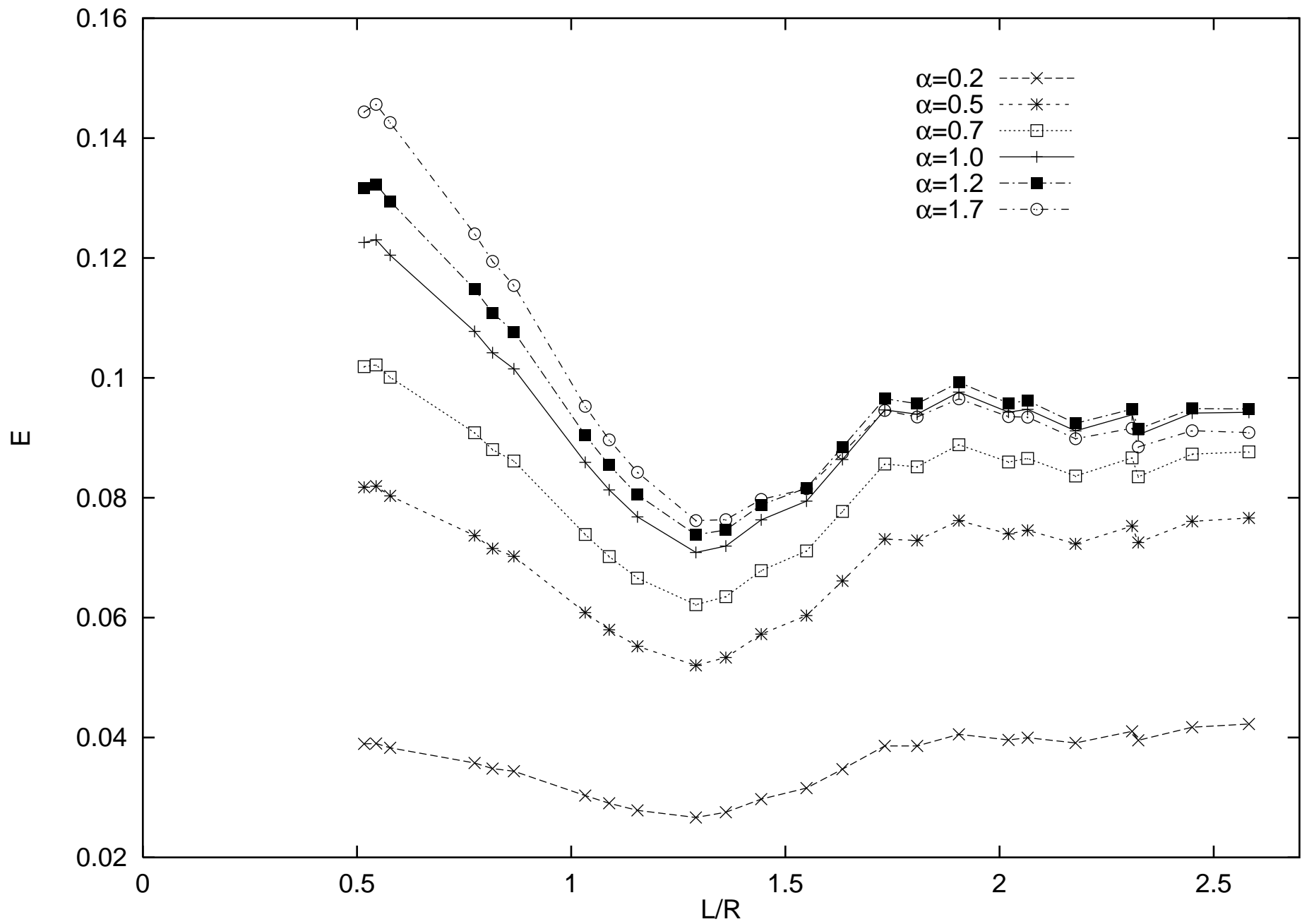
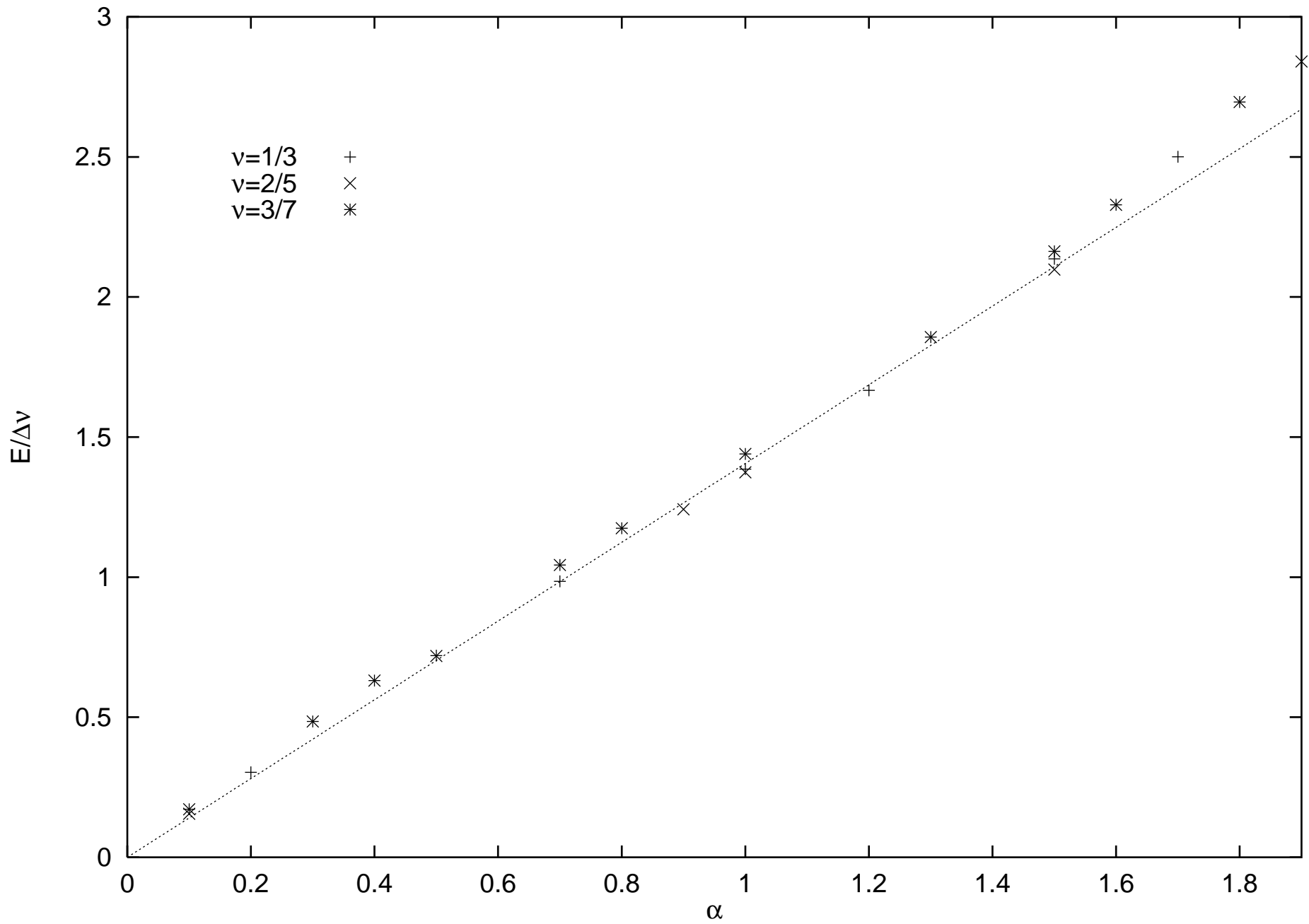
\end{document}